\documentstyle[aps,amssymb,eqsecnum]{revtex}
 \title{Noncommutative Martin-Lof randomness : on the concept of a random sequence of qubits}
 \author{Gavriel  Segre \thanks{AKNOWLEDGMENTS: I sincerely want to thank my true Teachers: prof. G. Jona-Lasinio and doct. F. Benatti , for their encouragement, for their teachings and for many precious suggestions and remarks about the issues discussed in this paper.They have no responsibility for any mistake contained in these pages.}}
 \address{Dipartimento di Fisica Nucleare e Teorica and I.N.F.N. - Universit\'{a} di Pavia}
 \address{Via Bassi 6 - 27100 Pavia, Italy}
 \newtheorem{definition}{DEF}[section]
 \newtheorem{theorem}{Theorem}[section]
 \newtheorem{trial definition}{Trial definition}[section]
 \newtheorem{remark}{Remark}[section]
 \newtheorem{axiom}{AXIOM}[section]
 \newenvironment{hypothesis} { HP: \vspace{1pt} \begin{center}} {\end{center}}
 \newenvironment{thesis} { TH: \vspace{1pt} \begin{center}} {\end{center}}
 \newtheorem{conjecture}{Conjecture}[section]
 \begin{document}
 \draft
 \maketitle
 \begin{abstract}
 Martin-Lof's definition of  random sequences of cbits as those not
 belonging to any set of constructive zero Lebesgue measure is reformulated
 in the language of Algebraic Probability Theory.

 The adoption of the Pour-El Richards theory of computability
 structures on Banach spaces allows us to give a natural
 noncommutative extension of Martin-Lof's definition ,
 characterizing the random elements of  a  chain Von Neumann
 algebra.

 In the particular case of  the  minimally informative
 noncommutative alphabet  our definition  reduces to the
 definition of a random sequence of qubits.
 \end{abstract}
 \section{Introduction} \label{Introduction}
 Probability Theory may be defined as that branch of Mathematics
 studying random phaenomena.

 The mathematical foundation of Classical Probability Theory was
 given by A.N. Kolmogorov in terms of Measure Theory : a classical
 probability space is nothing but a measure space in which the
 measure is normalized to one \cite{Billingsley 95}.

 As far as physical phaenomena are concerned, Quantum Physics ,
 differing in this by Classical Physics , is not deterministic:
 according to the Modern Statistical  Interpretation (that we adhere) quantum
 measurement is a random phaenomenon \cite{Holevo 99}.

 In spite of being formalizable in the language of Classical
 Probability Theory, quantum randomness presents some peculiarity
 that  led Feynman  to implicitely suggest that its deep comprehension
 requires the introduction of a Quantum Probability Theory
 different from the classical one:
 \begin{enumerate}
 \item the composition property  is satisfied by  classical -probability amplitudes and not by their
 square-moduli \cite{Feynman 48}
 \item  a generic quantum system can't be simulated efficentely by a classical universal computer \cite{Feynman 82}
 \end{enumerate}
 implying that the complexity class QP of  the problems soluble
 with certainty in the worst case in polynomial time by   a  (
 quantum-probabilistically - non-deterministic ) - computer is
 different both from the class P of  the problems  soluble  with
 certainty in the worst case in polynomial time by a deterministic
 classical computer and from the class NP of  the problems  soluble
 with certainty in the worst case in polynomial time by a
 non-deterministic classical computer \cite{Preskill 98},\cite{Williams Clearwater 98} ( i.e. quantum
 non-determinism is different both from classical determinism and from
 classical non-determinism, a thing unfortunately never
 considered in all the discussions about the possibility of a
 deterministic completion of quantum mechanics ).

 The adoption of the W*-algebraic language, in particular,  allows
 a unified formulation of both Classical and Quantum Physics , the
 classical case being characterized by the abelianity of the
 observables' algebra \cite{Thirring 81}, \cite{Thirring 83},
 \cite{Connes 94},  leading , through Noncommutative Measure Theory
 , to the delineation of Quantum Probability Theory as the
 noncommutative generalization of Classical Probability Theory.

 Now it is well known that , couriously,  Classical Probability
 Theory is not self-contained as far as the definition of  a random
 sequence on a finite alphabet is concerned : its appropriate
 characterization , given by Martin-Lof \cite{Martin-Lof 66a},
 \cite{Martin-Lof 66b}, \cite{Chaitin 87}, \cite{Calude 94}, requires notions from a
 field of Mathematics very far from Measure Theory, i.e. Classical
 Recursion Theory \cite{Odifreddi 89}.

 In this paper we analyze the same issue for Quantum Probability
 Theory: a brief review of Martin-Lof's randomness in
 sect.\ref{Martin-Lof random sequences} is reformulated, in
 sect.\ref{Martin-Lof randomness in the language of Von Neumann
 algebras} in the language of W*-algebras.

 The adoption of the Pour-El Richards theory of computability
 structures on Banach spaces allows us to give a natural
 noncommutative extension of Martin-Lof definition , characterizing
 the random elements of  a  chain Von Neumann algebra.

 In the particular case of the  minimally informative
 noncommutative alphabet  our definition  reduces to the
 definition of a random sequence of qubits.
 \section{Martin-Lof random sequences} \label{Martin-Lof random sequences}
 Given a finite alphabet $ \Sigma $  let us denote with $
 \Sigma^{\star} $ the set of the strings on $ \Sigma $ and with $
 \Sigma^{\infty} $ the set of the sequences on $ \Sigma $.Since:
 \begin{equation} \label{cardinality of the set of strings on a finite alphabet}
 card(\Sigma^{\star})=\aleph_{0}
 \end{equation}
 \begin{equation} \label{cardinality of the set of sequences on a finite alphabet}
 card(\Sigma^{\infty})=\aleph_{1}
 \end{equation}
 we can make, modulo recursive codings , the following
 identifications:
 \begin{equation}
 \Sigma \circeq \{0,1\}
 \end{equation}
 \begin{equation}
 \Sigma ^{*} \circeq \mathbb{N}
 \end{equation}
 \begin{equation}
 \Sigma^{\infty} \circeq ( 0 , 1 ]
 \end{equation}
 Denoted by $ { \mathcal{F} }_{cylinder} $ the $ \sigma $-algebra
 of the cylinder sets on $ \Sigma $ \cite{Billingsley 95}  let us
 consider a probability measure P on the measurable space $ (
 \Sigma^{\infty} , { \mathcal{F} }_{cylinder} ) $.

 The more natural way to characterize the random sequences on $
 \Sigma $ would seem to be the following: introduced the notion of:
 \begin{definition}[NOT-CONSTRUCTIVE P-TYPICAL SUBSETS
 OF $\Sigma^{\infty} $ ] \label{not-constructive P-typical
 sequences}
 \end{definition}
 \begin{equation}
 P-TYP(\Sigma)_{NC} \equiv \{ S \subseteq \Sigma^{\infty} : \neg (
 \forall \epsilon > 0 \exists \{ A_{n} \subset \Sigma^{\infty}
 \}_{n \in \mathbb{N}} : S \subset \bigcup_{ n \in N } A_{n} \wedge
 \sum_{n=0}^{\infty} P(A_{n}) < \epsilon ) \}
 \end{equation}
 it should be natural to call random those sequences on $ \Sigma $
 not belonging to any not-constructive P-typical set.

 Such an approach is, anyway, invalidated by the following:
 \begin{theorem}[on the not self-sufficiency of Measure Theory to define
 randomness] \label{on the not self-sufficiency of Measure Theory
 to define randomness}
 \end{theorem}
 \begin{equation}
 P-TYP(\Sigma)_{NC} = \emptyset \; \forall \; probability \;
 measure \; P \; on \; ( \Sigma^{\infty} , { \mathcal{F}
 }_{cylinder})
 \end{equation}
 The above approach was saved by P. Martin-Lof at the prize of
 requiring the introduction of ingredients  not belonging to Measure Theory  but to Classical
 Recursion Theory : introduced, in fact, the following notion:
 \begin{definition}[CONSTRUCTIVE P-TYPICAL  SUBSETS
 OF $ \Sigma^{\infty} $] \label{constructive P-typical sequences}
 \end{definition}
 \begin{equation}
 P-TYP(\Sigma)_{C} \equiv \{ S \subseteq \Sigma^{\infty} : \neg (
 \forall \; \epsilon > 0 \exists \{ A_{n} \subset \Sigma^{\infty}
 \}_{n \in \mathbb{N}} :\{ A_{n} \}\; is \; r.e. \wedge S \subset
 \bigcup_{ n \in N } A_{n} \wedge \sum_{n=0}^{\infty} P(A_{n}) <
 \epsilon ) \}
 \end{equation}
 depending on the recursion-theoretic notion of
 recursive-enumerability ( r.e.-bility ) and given a
 sequence $ \{ \omega_{n} \}_{n \in \mathbb{N}} $ on $ \Sigma $ we
 may characterize its randomness in the following way:
 \begin{definition} [ $ \{ \omega_{n} \}_{n \in { \mathbb{N} } } $
 is Martin-Lof random ]
 \end{definition}
 \begin{equation}
 \{ \omega_{n} \}_{n \in {\mathbb{N}}} \notin S \; \forall S \in
 P_{Lebesgue} - TYP ( \Sigma )_{C}
 \end{equation}
 where $ P_{Lebesgue} - TYP ( \Sigma )_{C} $ is the Lebesgue
 probability measure on $ \Sigma^{\infty} $.

 \section{Martin-Lof randomness in the language of Von Neumann algebras} \label{Martin-Lof randomness in the language of Von Neumann algebras}
 Demanding to the monograph of W.Thirring \cite{Thirring 81},
 \cite{Thirring 83} for any notion about Von Neumman algebras and
 their link with Quantum Mechanics, let us introduce the following
 notion:
 \begin{definition} [ ALGEBRAIC PROBABILITY SPACE ] $ ( A ,
 \omega ) $ :
 \end{definition}
 \begin{itemize}
 \item A is a W*-algebra
 \item $ \omega \in S(A) $ is a state on A
 \end{itemize}
 The notion of algebraic probability space is a noncommutative
 generalization of the notion of classical probability space,i.e.
 the concepts of classical probability space and abelian algebraic
 probability space are conceptually equivalent as is shown by the
 following facts:
 \begin{enumerate}
 \item given a classical probability space $ (S, {\mathcal{F}},P) $
 let us observe that it can also be described in a different,
 conceptually equivalent, way as the couple $ (A , \omega ) $
 where:
 \begin{itemize}
 \item A is the abelian W*-algebra $ L^{\infty} (S,P) $
 \item $ \omega_{P} \in S(A) $ is the state on A defined by:
 \end{itemize}
 \begin{equation} \label{state of a classical probability space}
 \omega_{P} (a) \equiv \int_{S} a(x) dP(x) \; a \in A
 \end{equation}
 \item given an abelian W*-algebra A and a state on it $ \omega \in
 S(A) $, there  exist a classical probability space $ (S,
 {\mathcal{F}},P) $ such that $ A = L^{\infty} (S,P) $ and $ \omega
 = \omega_{P} $.
 \end{enumerate}
 Let us, then, reformulate the notion of Martin-Lof randomness in
 the algebraic language substituing to the classical probability
 space $ (S , { \mathcal{F} }_{cylinder} , P ) $ the corrispective
 abelian algebraic probability space $ ( L^{\infty}(S,P),\omega_{P}
 ) $; then to a subset $ A \subset \Sigma^{\infty} $ corresponds
 its characteristic function $ \chi_{A} \in L^{\infty}(
 \Sigma^{\infty},P) $.

 Instead of the P-typical subsets of $ \Sigma^{\infty} $ we must,
 then, look at the $ \omega_{P} $-typical subsets of $ L^{\infty}(
 \Sigma^{\infty},P) $  whose definition may be easily obtained
 traducing the definitions DEF \ref{not-constructive P-typical
 sequences} and DEF  \ref{constructive P-typical sequences} in
 terms of functions on $ \Sigma^{\infty} $:
 \begin{definition}[NOT-CONSTRUCTIVE $
 \omega_{P}$-TYPICAL SETS OF  $ L^{\infty}( \Sigma^{\infty},P) $]
 \label{not-constructive P-typical sequences in W*-algebraic
 language}
 \end{definition}
 \begin{equation}
 \omega_{P} - TYP [ L^{\infty}( \Sigma^{\infty},P) ]_{NC} \equiv \{
 S \subseteq  L^{\infty}( \Sigma^{\infty},P) : \neg ( \forall
 \epsilon > 0 \exists  \{ A_{n}\}  \subset \Sigma^{\infty} :
 eq.\ref{C1.2} \wedge eq.\ref{C1.3} ) \}
 \end{equation}
 where:
 \begin{equation} \label{C1.2}
 S \subset span ( \{ \chi_{A_{n}} \}_{n \in {\mathbb{N}}} )
 \end{equation}
 \begin{equation} \label{C1.3}
 \sum_{n=0}^{\infty} \omega_{P} ( A_{n} ) < \epsilon
 \end{equation}
 \begin{definition}[CONSTRUCTIVE $
 \omega_{P}$-TYPICAL SETS OF  $ L^{\infty}( \Sigma^{\infty},P) $]
 \label{constructive P-typical sequences in W*-algebraic language}
 \end{definition}
 \begin{equation}
 \omega_{P} - TYP [ L^{\infty}( \Sigma^{\infty},P) ]_{C} \equiv \{
 S \subseteq  L^{\infty}( \Sigma^{\infty},P) : \neg ( \forall
 \epsilon > 0 \exists \{ A_{n} \} \subset \Sigma^{\infty} :
 eq.\ref{C1.1} \wedge eq.\ref{C1.2} \wedge eq.\ref{C1.3} ) \}
 \end{equation}
 where:
 \begin{equation} \label{C1.1}
 \{ \chi_{A_{n}} \}_{n \in {\mathbb{N}}} \; is \; recursively \;
 enumerable
 \end{equation}
 and where $ span ( \{ \chi_{A_{n}} \}_{n \in {\mathbb{N}}} ) $ is
 the W*-subalgebra of $ L^{\infty} ( \Sigma^{\infty} , P ) $
 generated by the characteristic functions $ \{ \chi_{A_{n}} \}_{n
 \in {\mathbb{N}}} $.

 The necessity of introducing the constructive $ \omega_{P}
 $-typical sets of  $ L^{\infty} ( \Sigma^{\infty} , P ) $ lies,
 obviously, again, on the theorem \ref{on the not self-sufficiency
 of Measure Theory to define randomness} whose traduction in
 algebraic form is the following:
 \begin{theorem}[W*-algebraic formulation of theorem \ref{on the not self-sufficiency of Measure Theory
 to define randomness}]
 \end{theorem}
 \begin{equation}
 \omega_{P}-TYP[L^{\infty}(\Sigma^{\infty},P)])_{NC} = \emptyset \;
 \forall \; probability \; measure \; P \; on \; ( \Sigma^{\infty}
 , { \mathcal{F} }_{cylinder})
 \end{equation}
 Given a function $ f \in L^{\infty}(\Sigma^{\infty},P)$ we have,
 then , clearly, that:
 \begin{theorem} \label{W*-algebraic condition equivalent to Martin-Lof randomness}
 \end{theorem}
 \begin{equation}
 f \; is \; Martin-Lof \; random \Leftrightarrow f \notin S \; \forall S
 \in
 \omega_{P_{Lebesgue}}-TYP[L^{\infty}(\Sigma^{\infty},P)]_{C}
 \end{equation}
 As far as the definitions  DEF\ref{not-constructive P-typical
 sequences} and DEF\ref{constructive P-typical sequences} are concerned
 we have already remarked  that they involve , through the
 constructivity assumption, a field  of Mathematics very far from
 Measure Theory, i.e. Recursion Theory.

 More precisely they involve that standard part of Recursion Theory
 that, through robust theorems of equivalence between
 different approaches,is  univoquely determined,i.e. Classical
 Recursion Theory \cite{Odifreddi 89}.

 Such a lucky situation doesn't occur  , as we will see in the next
 paragraph, when their noncommutative extensions are considered.
 \section{Noncommutative constructive measure theory and randomness} \label{Noncommutative constructive measure theory and randomness}
 Condensed in a slogan the leit-motif of Noncommutative Geometry ,
 corroborated by the observation presented in the previuos
 paragraph, consists in  looking at an algebraic probability space $
 ( A , \omega ) $  as '' a kind of '' $ (
 L^{\infty}(S,P),\omega_{P} ) $ with S a '' noncommutative space ''
 and P a noncommutative probability measure on S \cite{Connes 94}.

 Looking at the definition DEF \ref{constructive P-typical sequences in W*-algebraic
 language} the issue of looking for an its natural noncommutative extension
 would appear as a typical research program of Noncommutative
 Geometry .

 Such a definition , and the conseguent characterization of
 Martin-Lof randomness given by Theorem \ref{W*-algebraic condition
 equivalent to Martin-Lof randomness}  anyway, involving Recursion
 Theory, belongs to a  peculiar , surprisingly unexlored ,field
 of research: \textbf{Noncommutative Constructive Measure theory}.

 Given an algebraic probability space $ ( A , \omega ) $ a naive
 extension of the definition DEF: \ref{constructive P-typical
 sequences in W*-algebraic language} would sound as follow :
 \begin{trial definition}[CONSTRUCTIVE $\omega$-TYPICAL SETS OF
 A] \label{trial definition of typical subset of a W*-algebra
 w.r.t. a state}
 \end{trial definition}
 \begin{equation}
 \omega -TYP[A]_{C} \equiv \{ B \subseteq A : \neg ( \forall
 \epsilon > 0 \exists \{ a_{n} \in A \}_{n \in {\mathbb{N} }} :
 eq.\ref{C2.1} \wedge eq.\ref{C2.2} \wedge eq.\ref{C2.3} \}
 \end{equation}
 where:
 \begin{equation} \label{C2.1}
 \{ a_{n}\} \; is \; recursively \; enumerable
 \end{equation}
 \begin{equation} \label{C2.2}
 B \subset span( \{ a_{n} \}_{n \in {\mathbb{N}}} )
 \end{equation}
 \begin{equation} \label{C2.3}
 \sum_{n=0}^{\infty} \omega ( a_{n}^{*} a_{n} ) < \epsilon
 \end{equation}
 Granted, for the moment, that such a definition is correct, it leads
 immediately to the following definition:
 \begin{trial definition}[ $ a \in A $ is random ] \label{trial definition of noncommutative randomness}
 \end{trial definition}
 \begin{equation}
 a \notin B \; \forall B \in \omega_{Lebesgue}-TYP[A]_{C}
 \end{equation}
 The definitions DEF:\ref{trial definition of typical subset of a W*-algebra w.r.t. a state} and DEF:\ref{trial definition of noncommutative randomness}
 are vague and imprecise owed to the
 following:
 \begin{remark} \label{not-existence of a natural noncommutative generalization of recursive enumerability}
 it doesn't exist, in Recursion Theory,  a standard definition of a
 recursively-enumerable sequence on a Von Neumann algebra
 \end{remark}
 \begin{remark} \label{not-existence of a Lebesgue state on a W*-algebra}
 it doesn't exist any notion of a Lebesgue-state on a W*-algebra
 \end{remark}

 As far as REMARK \ref{not-existence of a natural noncommutative
 generalization of recursive enumerability}  is concerned  a
 particularly promizing ,though not-univoquely determined,
 extension of Computable Analysis to the theory of Banach spaces
 has been realized by   M. Pour-El and J.I. Richards
 \cite{Pour-El Richards 89}.

 Given a Banach space B  on the real/complex field Pour-El and
 Richards introduce the following notion:
 \begin{definition} [COMPUTABILITY STRUCTURE on B] \label{computability structure on a Banach space}
 a specification of a subset $ {\mathcal{S}} $ of the set of all
 the sequences  in B identified as the \textbf{set of the
 computable sequences on B} :
 \end{definition}
 \begin{axiom}[linear forms] \label{linear forms}
 \end{axiom}
 \begin{hypothesis}
 $ \{ x_{n} \} $ and  $ \{ y_{n} \} $ computable sequences in B

 $ \{ \alpha_{n,k} \} ,  \{ \beta_{n,k} \} $ two recursive double
 sequence of real/complex numbers

 d recursive function

 $ s_{n} \equiv \sum_{k=0}^{d(n)} \alpha_{n,k} x_{k} + \beta_{n,k}
 y_{k} $
 \end{hypothesis}
 \begin{thesis}
 $ \{ s_{n} \} \in { \mathcal{S} } $
 \end{thesis}
 \begin{axiom}[limits] \label{limits}
 \end{axiom}
 \begin{hypothesis}
 $ x_{n,k} $ computable double sequence in B : $ r- \lim_{k
 \rightarrow \infty} x_{n,k} =  x_{n}  $
 \end{hypothesis}
 \begin{thesis}
 $ \{ x_{n} \} \in {\mathcal{S}} $
 \end{thesis}
 \begin{axiom}[norms] \label{norms}
 \end{axiom}
 \begin{hypothesis}
 $ \{ x_{n} \} \in {\mathcal{S}} $
 \end{hypothesis}
 \begin{thesis}
 $ \{ \| x_{n} \| \} $ is a  recursive sequence of real  numbers.
 \end{thesis}
 These axioms contains some notion we have to specify.

 First of all let us observe that, given a sequence  $ \{ x_{n} \}
 $ of real or complex numbers, the fact that each element  of the
 sequence is recursive, and can, consequentely, be effectively
 approximated to any desired degree of  precision  by a computer
 program $ P_{n} $ given in advance doesn't imply the recursivity
 of the whole sequence since there might not exist a way of
 combining the sequence of programs $ \{ P_{n} \} $  in a unique
 program P computing the whole sequence $ \{ x_{n} \} $.

 Given a double sequence  $ \{ x_{n,k} \in \mathbb{R} \} $ and an
 other sequence  $ \{ x_{n} \} $ of real numbers :
 \begin{equation}
 \lim_{ k \rightarrow \infty }  x_{n,k} = x_{n} \;  \forall n \in
 \mathbb{N}
 \end{equation}
 \begin{definition} [ $ \{ x_{n,k} \} $ CONVERGES RECURSIVELY TO $ \{ x_{n} \} ( r- \lim_{k \rightarrow \infty } x_{n,k} = x_{n}
 ) $ ]
 \end{definition}
 \begin{equation}
 \exists e: { \mathbb{N} } \times { \mathbb{N} } \rightarrow {
 \mathbb{N} } \in \Delta_{0}^{0} : ( k > e(n,N) \Rightarrow \mid
 r_{k} - x \mid \leq  \frac{1}{2^{N}} ) \; \forall n \in { \mathbb{N}
 } , \; \forall N \in { \mathbb{N} }
 \end{equation}
 \begin{definition} [  $ \{ x_{n} \}_{n \in {\mathbb{N}}} $ IS
 RECURSIVE ]
 \end{definition}
 \begin{equation}
 \exists \{ r_{n,k} \in { \mathbb{Q} } \}_{n,k \in { \mathbb{N} }
 } : \mid r_{n,k} - x_{n} \mid \leq \frac{1}{2^{k}}
 \end{equation}
 \begin{definition} [ $ \{ z_{n} \in {\mathbb{C}} \}_{n \in
 {\mathbb{N}}} $ ]
 \end{definition}
 \begin{equation}
  \{ \Re (z_{n} ) \}_{n \in {\mathbb{N}}} \; and \; \{ \Im (z_{n}
 ) \}_{n \in {\mathbb{N}}} \; are \; recursive
 \end{equation}

 The above argument should clarify why the definition of a
 computability structure on a Banach space B is made through a
 proper specification of  the computable sequences in B and not,
 simply, by  the specification of a proper set of  the computables
 vectors.

 The notion of a computable vector, instead, is immediately induced
 by the assignment on  B of a computability structure $ \mathcal{S}
 $ .
 \begin{definition}[COMPUTABLE VECTORS OF B]
 \end{definition}
 \begin{equation}
 \Delta_{0}^{0} (B) \equiv \{ x \in B : \{x,x,x, \ldots \} \in
 \mathcal{S} \}
 \end{equation}

 The Axioms Axiom\ref{linear forms}, Axiom\ref{limits} and Axiom\ref{norms} have a
 transparent intuitive meaning: since a Banach space is  made up of
 a vector space V, a norm on V and the completeness-condition for such a norm, it is natural to require analogous effective conditions
 for the set of computable sequences.

 Unfortunately  such axioms do not provide the axiomatic definition
 of a unique structure for a Banach  space B  since B admits,
 generally, more computability-structures.

 This, anyway, doesn't relativize the whole approach thanks to the
 existence of a suppletive condition whose satisfability results in
 the invoked univocity.

 Given a computability structure  $ \mathcal{S} $  on a Banach
 space B:
 \begin{definition}[EFFECTIVE GENERATING SET for B]
 \end{definition}

 \begin{equation}
  \{ e_{n} \} \in {\mathcal{S}} \; : \; linear-span( \{ e_{n} \} )
  is \; dense\; in \; B
 \end{equation}
 \begin{definition}[B is EFFECTIVELY SEPARABLE]
 \end{definition}
 \begin{equation}
 \exists \{ e_{n} \} \; effective \; generating \; set \; for \;
 B
 \end{equation}
 Fortunately Pour-El and Richards  proved the following:
 \begin{theorem}[of stability]
 \end{theorem}
 \begin{hypothesis}
 B Banach space

 $ {\mathcal{S}}_{1} $ , $ {\mathcal{S}}_{2} $  effectively separable computability structures on B

 $ \{ e_{n} \} \in {\mathcal{S}}_{1} \cap {\mathcal{S}}_{2} $
 effective generating set for B
 \end{hypothesis}
 \begin{thesis}
 $ {\mathcal{S}}_{1} = {\mathcal{S}}_{2} $
 \end{thesis}
 Let us now return to our algebraic probability space $ ( A , \omega
 ) $ and let us suppose that A is endowed with a computability
 structure $ {\mathcal{S}} $ eventually associated to some effective
 generating set of  observables physically known to be effectively
 measurable.

 Then it would  appear meaningful  to  give meaning to the  vague
 locution \emph{ '' $ \{ a_{n} \in A \}_{n \in {\mathbb{N}}} $ is
 recursively-enumerable '' } in eq.\ref{trial definition of
 typical subset of a W*-algebra w.r.t. a state}   substituing it
 with the precise condition \emph{ '' $ \{ a_{n} \in A \}_{n \in
 {\mathbb{N}}}
  \in {\mathcal{S}} $ '' }.

 As far as Remark\ref{not-existence of a Lebesgue state on a W*-algebra}
 is concerned, let us observe that $ P_{Lebesgue} $,  being the ''
 uniform probability measure '' on $ \Sigma^{\infty} $, is that of
 maximum entropy.

 Then it appears natural to think that the role of the unprecised state  $
 \omega_{Lebesgue} $ in \ref{trial definition of noncommutative randomness}  must be played by a  state on A of maximum entropy according to the following definition :

 \begin{definition}[ENTROPY of a state $ \omega $ on A ]
 \end{definition}
 \begin{equation}
 S ( \omega ) \equiv \{ \sum_{i} \lambda_{i} S(\omega_{i} , \, \omega )  : \sum_{i} \lambda_{i} \omega_{i} = \omega \}
 \end{equation}
 where the supremum is taken over all the decompositions of $
 \omega $ in countable convex combinations of other states and
 where $ S(\omega_{i} \, , \, \omega ) $  is the Umegaki-Araki
 relative entropy of $ \omega_{i} \; w.r.t. \; \omega $ \cite{Ohya
 Petz 93}.

 Collecting our considerations about Remark\ref{not-existence of a natural noncommutative generalization of recursive enumerability} and Reamrk\ref{not-existence of a Lebesgue state on a W*-algebra} we are, then, lead to introduce the following definitions:
 \begin{definition}[CONSTRUCTIVE $ \omega $-TYPICAL SETS OF $ (A ,
 {\mathcal{S}} ) $ ]
 \end{definition}
 \begin{equation}
  \omega -TYP[A]_{C} \equiv \{ B \subseteq A : \neg ( \forall
 \epsilon > 0 \exists \{ a_{n} \in A \}_{n \in {\mathbb{N} }} :
 eq.\ref{C3.1} \wedge eq.\ref{C2.2} \wedge eq.\ref{C2.3} \}
 \end{equation}

 where:

 \begin{equation} \label{C3.1}
 \{ a_{n} \}_{n \in {\mathbb{N}}} \in \mathcal{S}
 \end{equation}
 \begin{definition}[$ a \in {\mathcal{S}}$ is random]  \label{noncommutative randomness}
 \end{definition}
 \begin{equation}
 a \notin B  \; \forall B \in \omega_{Maximum \; Entropy} - TYP [A]_{C}
 \end{equation}

 where $ \omega_{Maximum \: Entropy} $ is the state on A  of maximum entropy  provided such a state exists.

 Though the definition  DEF\ref{noncommutative randomness} was given for an arbitary W*-algebra endowed with a computability structure , its physical relevance arises when the particular case of chain-algebras is considered:

 let us observe, in fact, that , in the previous paragraph, we
 didn't consider arbitrary probability spaces but  the  particular
 probability spaces $ ( \Sigma^{\infty} , { \mathcal{F}
 }_{cylinder} , P ) $  on the sequences on  a finite alphabet $
 \Sigma $; the physically relevant applications of the definition
 \ref{noncommutative randomness}  correspond, then,  to the case
 of algebraic probability spaces on the set  of the sequences on a \emph{'' noncommutative alphabet '' }.

 Such particular algebraic probability spaces are  nothing but the
 well known  chain  - quasi-local - algebras of the one-dimensional quantum lattice spins systems usual
 in Quantum Statistical Mechanics \cite{Ohya Petz 93}.

 As far as usual Martin-Lof randomness is concerned the relations
 \ref{cardinality of the set of strings on a finite alphabet} and
 \ref{cardinality of the set of sequences on a finite alphabet}
 explain why  all finite alphabets are, as far as the definition of
 randomness is concerned, absolutely equivalent and justifies ,
 conseguentely, the reduction of the analysis to the  alphabet with
 minimal classical information, i.e to the one cbit alphabet $
 \Sigma \circeq \{0,1\} $.

 In the same way, we can, as far as the definition of a random
 noncommutative sequence is concerned, restrict the analysis to
 the alphabet of minimal \emph{ '' noncommutative information '' },
 i.e. to the 1-qubit alphabet $ \{0,1\}_{NONCOMMUTATIVE} $ with
 observable algebra \cite{Preskill 98} :

 \begin{equation}
 M_{2} ( { \mathbb{C} } ) \equiv \{ \vec{\alpha} \cdot \vec{\sigma}
 + \beta {\mathbb{I}}, \; \vec{\alpha} \in {\mathbb{C}}^{3} , \beta
 \in {\mathbb{C}} \}
 \end{equation}
where $ \vec{\sigma}  \equiv ( \sigma_{1} , \sigma_{2} ,
 \sigma_{3} ) $ is the 3-vector of Pauli matrices and $ \mathbb{I} $ is the bidimensional identity matrix.

 Since both the measurements of spin-components and the
 preservation of the state are effective operations, it appears
 physically reasonable to consider A endowed with the
 computability structure $ \mathcal{S} $ individuated by the condition that the basis of A:

 \begin{equation}
 {\mathcal{S}}_{spin} \equiv \{ \sigma_{1} , \sigma_{2} ,
 \sigma_{3} , {\mathbb{I}} \}
 \end{equation}

 is effective.

 The computable matrices in $ M_{2} ( { \mathbb{C} } ) $ are, informally speaking, those  obtainable by elements of $ {\mathcal{S}}_{spin} $ with '' effective linear combinations ''.

 As an example let us consider  the following matrices:
 \begin{equation}
 m_{Collatz} ( \vec{n} ) \equiv \frac{1}{2} ( 1 + ( - 1
 )^{p_{Collatz} ( | \vec{n} |+1 ) } \vec{n} \cdot \vec{\sigma} ) \;
 \vec{n} \in {\mathbb{N}}^{3}
 \end{equation}
 where $ p_{Collatz} : {\mathbb{N}} \rightarrow \{ 0 ,1 \} $ is the Collatz predicate:
 \begin{equation}
 p_{Collatz} (n) = ( cond \; \exists m \in {\mathbb{N}} \; : \; T^{m} (n) = 1 \;,\; 1 \; , \; 0 )
 \end{equation}
 while $   T^{m} (n) $ is the sequence:
 \begin{equation}
  T^{m} (n) \equiv ( cond \; m = 1 \; , \; n \; , \; ( \; cond \; T^{m-1} (n) \; is \; even\;
 \; , \; \frac{  T^{m-1} (n)}{2} \; , \; 3 T^{m-1} (n) ))
 \end{equation}
 where I have adopted Mc Carthy's LISP notation for the conditional
 definitions \cite{Mc Carthy 60}. The well-known
 recursive-undecidability \cite{Conway 72} of the:
 \begin{conjecture}[of Collatz]
 \end{conjecture}
 \begin{equation}
 p_{Collatz}(n) = 1 \; \forall n \in
 {\mathbb{N}}
 \end{equation}
 implies the recursive-undecidability of the associated statement
 in term of  the matrices $ m_{Collatz} ( \vec{n} ) $ :
 \begin{conjecture}[ of Collatz in terms of the W*-algebra $ M_{2}({\mathbb{C}}) $ ] $ m_{Collatz} ( \vec{n} ) $ is the projector
 associated to spin +1/2  along the component $ \vec{n} \; , \;
 \forall \vec{n} \in {\mathbb{N}}^{3} $
 \end{conjecture}
 The matrices $ m_{Collatz} ( \vec{n} ) $ are given by a
 non-effective linear combination of the elements of
 $ {\mathcal{S}}_{spin} $ and are, then, non computable.

 Let us , then, consider the 1-qubit spin chain:
 \begin{equation}
 A \equiv norm-completion ( A_{ { \mathbb{Z}}} ) \; , \;  A_{ {
 \mathbb{Z}}} \equiv \bigotimes_{{\mathbb{Z}}} M_{2}( {\mathbb{C}} )
 \end{equation}
 A  state of maximum entropy on A is, clearly, that represented by the density matrix:
 \begin{equation}
 \rho_{uniform} \equiv \times_{{\mathbb{Z}}} \;
 diagonal(\frac{1}{2},\frac{1}{2})
 \end{equation}
 describing  the thermodinamical equilibrium of a  one-dimensional lattice of spins-1/2 at infinite temperature.

 Furthermore, the computability structure $ {\mathcal{S}}_{spin} $
 induces, naturally, the  computability structure on A:
 \begin{equation}
 {\mathcal{S}} \equiv \times_{{\mathbb{Z}}} {\mathcal{S}}_{spin}
 \end{equation}
 Given ,then, a sequence of qubits $ a \in A $, i.e. an
 infinitely long word on the alphabet $ \{0,1\}_{NONCOMMUTATIVE} $ of minimal noncommutative information, we have , as a particular case of the general definition DEF\ref{noncommutative randomness}, that:
 \begin{equation}
 a \; is \; random \; \Leftrightarrow \; a \notin B \; \forall B
 \in \omega_{uniform}-TYP[A]_{C}
 \end{equation}
 where $ \omega_{uniform}-TYP[A]_{C} $ is the set of all the
 constructive $ \omega_{uniform} $ - typical sets  of $ ( A ,
 {\mathcal{S}}_{chain} ) $,i.e.:
 \begin{equation}
 \omega -TYP[A]_{C} \equiv \{ B \subseteq A : \neg ( \forall
 \epsilon > 0 \exists \{ a_{n} \in A \}_{n \in {\mathbb{N} }} :
 eq.\ref{C4.1} \wedge eq.\ref{C2.2} \wedge eq.\ref{C4.3} \}
 \end{equation}
 with:
 \begin{equation} \label{C4.1}
 \{ a_{n} \}_{n \in {\mathbb{N}}} \in {\mathcal{S}}_{chain}
 \end{equation}
 \begin{equation} \label{C4.3}
 \sum_{n=0}^{\infty} \omega_{uniform} ( a_{n}^{*} a_{n} ) < \epsilon
 \end{equation}

 \end{document}